\documentclass[prd,aps,showpacs,floats,floatfix,eqsecnum]{revtex4}
\usepackage[dvips]{graphicx}
\usepackage{color}
\usepackage{subfigure}
\usepackage{amssymb}
\begin{document}
\title{ Compact stars in
Eddington-inspired Born-Infeld gravity: Anomalies associated with
phase transitions }
\author{Y.-H. Sham, P.~T. Leung\footnote{ptleung@phy.cuhk.edu.hk}, and L.-M. Lin}
\affiliation{Department of Physics and
Institute of Theoretical Physics, The Chinese University of Hong
Kong, Hong Kong SAR, China}
\begin{abstract}
  We study  how generic phase
transitions taking place in compact stars constructed in the
framework of the Eddington-inspired Born-Infeld (EiBI) gravity can
lead to anomalous behavior of these stars. For the
case with first-order phase transitions, compact stars in EiBI
gravity with a positive coupling parameter $\kappa $ exhibit a
finite  region with constant pressure, which is absent in general
relativity. However, for the case with a negative $\kappa $,
an equilibrium stellar configuration cannot be constructed. Hence,
EiBI gravity seems to impose stricter constraints on the
microphysics of stellar matter. Besides, in the presence of
spatial discontinuities in the sound speed $c_s$ due to phase
transitions, the Ricci scalar is spatially discontinuous and
contains $\delta$-function singularities proportional to the jump
in $c_s^2$ acquired in the associated phase transition.

\end{abstract}

%% 04.50.-h: Other theories of gravity
%% 04.40.Dg: Relativistic stars
%% 98.80.-k: Cosmology
\maketitle
\section{Introduction}\label{S1}
%The theory of general relativity (GR), founded in 1916 by
%Einstein, relates the curvature of spacetime to the
%motion/distribution of matter and energy and has sailed through
%numerous theoretical and experimental examinations, including the
%perihelion advance of Mercury, the deflection of light,  the
%orbital decay of  binary pulsars and the discoveries of blackholes
%(see~\cite{Will2006} for a review on experimental tests of GR).
%Notwithstanding the fact that GR has stood the test of time for
%nearly one century, other alternative theories of gravity
%(including scalar-tensor theories, $f(R)$ theories and
%higher-curvature gravity) have been proposed in attempt to  extend
%(or remedy) GR to certain extreme situations, such as strong
%gravity field, quantum gravity, and cosmological structure (see
%\cite{Clifton2012,Pani_alter} and references therein). For
%example, the observed accelerating expansion of the Universe has
%acted as a stimulus to the formulation of alternative theories of
%gravity (see, e.g., \cite{Clifton2012,Pani_alter,dark}).

Based on the gravitational action and nonlinear electrodynamics
originally proposed by Eddington \cite{Edd}, and Born and Infeld
\cite{Born}, respectively, Ba\~{n}ados and Ferreira recently
established a new Eddington-inspired Born-Infeld (EiBI) theory of
gravity \cite{Banados10}, which reduces to general relativity (GR)
in vacuum (see, e.g., \cite{Deser,Vollick} for relevant studies).
However, in the presence of coupling between matter, EiBI theory
demonstrates several distinctive features, including the avoidance
of singularities in the early cosmology and in the Newtonian
collapse of pressureless particles, the formation of stable
pressureless stars, and the existence of pressureless cold dark
matter with a nonzero Jeans length
\cite{Banados10,Pani11,Pani12,cosmo_constraint,Cho_12,bounce,
Sham12,Avelino12}. All such features stem from the effective
repulsive gravitational effect intrinsic to  EiBI  theory with a
positive coupling constant $\kappa$ (see the following discussion
for the definition of $\kappa$). Hence, in comparison with  GR,
EiBI theory also allows compact stars to attain higher mass limits
\cite{Pani11,Sham12}.

Unlike GR, in EiBI theory gravity and matter are nonlinearly
coupled together. Therefore, the distinctions between GR and EiBI
theory are expected to emerge in the high-density regime. To this
end, the static structure and the stability of compact stars in
EiBI theory have become the focus of several recent papers
\cite{Pani11,Pani12,Sham12}. As noted in these papers, the
presence of terms proportional to the spatial derivatives of the
density in the equations governing the hydrostatic
equilibrium and radial oscillation of compact stars in EiBI theory
could potentially hamper the related studies if nonanalytic (i.e., 
discontinuous or nondifferentiable) equations of state (EOSs) are
considered. To bypass the problems arising from such derivatives,
in the above-mentioned studies the EOSs employed in constructing
compact stars in EiBI theory are either polytropic or analytic
fits mimicking the behavior of realistic nuclear matter
\cite{Haensel04}. However, even for polytropic EOSs of the form $P
\propto \rho^\Gamma$, with $P$,   $\rho$, and $\Gamma$ being the
pressure,  rest-mass density, and a fixed constant, respectively,
Pani and Sotiriou have recently discovered that the Ricci scalar
diverges at the stellar surface if the constant $\Gamma$ is
greater than 3/2 \cite{Pani_sing}. Such singular behavior is
attributable to the auxiliary field $q_{\mu \nu}$ proposed  in
EiBI gravity. In the process of eliminating the auxiliary field in
the equation of motion, higher derivatives of the matter field
emerge and hence lead to a sensitive dependence on the matter
distribution.

In the present paper we investigate how the static structure of
compact stars in EiBI theory could be affected by drastic changes
in the EOS. Specifically, we consider  nuclear EOSs with
first-order (second-order) phase transitions, where the energy
density (the sound speed) is a discontinuous function of pressure.
Even in GR, the effects of such discontinuities are known to be
nontrivial \cite{Banik03,Sotani01}, and could lead to a third
family of compact stars \cite{third_1,third_2}. For compact stars
in EiBI gravity, we find that in the presence of first-order phase
transitions, no equilibrium configuration exists if the parameter
$\kappa < 0$. Furthermore, the Ricci scalar develops a
$\delta$-function singularity and discontinuity at the radius
where the speed of sound is discontinuous due to  phase
transitions. We note that such behavior is in stark contrast to
the case in GR.

The plan of the paper is as follows. We briefly summarize EiBI
gravity and the equilibrium configuration of compact stars in
Secs.~\ref{S2} and \ref{S3}, respectively.  In Sec.~\ref{S4} we
consider equilibrium stars in EiBI gravity constructed with EOSs
with first-order phase transitions and investigate the effects and
physical implications of such transitions.  The singular behavior
revealed  in the Ricci scalar as a consequence of the
discontinuity in the sound speed is studied in Sec.~\ref{S5}.
Finally, our conclusions are summarized in Sec.~\ref{S6}. We use
units where $G=c=1$ unless otherwise noted.

%%%%%%%%%%%%%%%%%%%%%%%%%%%%
\section{Eddington-inspired Born-Infeld gravity}\label{S2}
%%%%%%%%%%%%%%%%%%%%%%%%%%
Based on the Eddington gravity action and the Born-Infeld
nonlinear electrodynamics \cite{Edd,Born}, Ba\~{n}ados and
Ferreira established the EiBI gravity \cite{Banados10}, which is
summarized here as follows. The starting point of the  EiBI theory
is the action $S$ given by
\begin{eqnarray}
S &=& {1 \over 16 \pi} {2 \over \kappa} \int d^4x \left( \sqrt{
\left| g_{\mu\nu} + \kappa R_{\mu\nu} \right| } - \lambda \sqrt{-
g} \right)  \cr
&&\cr
&& + S_M \left[ g, \Psi_M \right] .
\label{eq:EiBI_action}
\end{eqnarray}
Here $R_{\mu\nu}$ is the symmetric part of the Ricci tensor and is
constructed solely from the connection
$\Gamma^\alpha_{\beta\gamma}$. However,
 the matter action $S_M$ depends only on the metric $g_{\mu\nu}$ and the matter field $\Psi_M$.
 Besides,  the determinant of a tensor $f_{\mu\nu}$ is denoted by
$\left| f_{\mu\nu} \right|$,  and the convention $f \equiv \left|
f_{\mu\nu} \right|$ is used hereafter. The parameters $\kappa$ and
$\lambda$ are related to the cosmological constant by $\Lambda =
(\lambda - 1) / \kappa$ such that, in the limit  $\kappa
\rightarrow 0$, the action (\ref{eq:EiBI_action}) reduces to the
Einstein-Hilbert action. In the present paper we aim at asymptotic
flat solutions and hence consider the case with $\lambda=1$.
Several constraints on the value and the sign of the remaining
parameter $\kappa$ have been obtained from solar observations, big
bang nucleosynthesis, and the existence of neutron stars
 \cite{Banados10,Pani11,Casanellas12,Avelino12}. In particular,
 for cases with positive $\kappa$, effective gravitational
 repulsion prevails, leading to the existence of pressureless
 stars and  increases in the mass limits of compact stars \cite{Pani11,Sham12}.

It is important to note that the spacetime metric $g_{\mu\nu}$ and
the connection $\Gamma^\alpha_{\beta \gamma}$ are treated as
independent fields in EiBI theory. The field equations are derived
by minimizing the action (\ref{eq:EiBI_action}) with respect to
$g_{\mu\nu}$ and $\Gamma^\alpha_{\beta\gamma}$ separately, and
read as \cite{Banados10}
\begin{eqnarray}
q_{\mu\nu} &=& g_{\mu\nu} + \kappa R_{\mu\nu} , \label{eq:field1} \\
\cr \sqrt{-q} q^{\mu\nu} &=&  \sqrt{-g} g^{\mu\nu} - 8\pi \kappa
\sqrt{-g} T^{\mu\nu} , \label{eq:field2}
\end{eqnarray}
with $q_{\mu\nu}$ being an auxiliary metric related to the
connection,
\begin{equation}\label{connection}
\Gamma^\alpha_{\beta\gamma} = {1 \over 2} q^{\alpha\sigma} \left(
\partial_{\gamma} q_{\sigma\beta} +\partial_{\beta}
q_{\sigma\gamma} - \partial_{\sigma} q_{\beta\gamma} \right) .
\end{equation}
On the other hand, the stress-energy tensor $T^{\mu\nu}$ still
satisfies the standard conservation equations,
\begin{equation}
\nabla_\mu T^{\mu\nu} = 0 , \label{eq:matter_conserv}
\end{equation}
where,  as in GR, the covariant derivative refers to the metric
$g_{\mu\nu}$.

It is worthwhile to note that both $R_{\mu\nu}$ and $T_{\mu\nu}$
can  lead to a difference in the two metrics $g_{\mu\nu}$ and
$q_{\mu\nu}$. If somehow there are some discontinuities in
$T_{\mu\nu}$ (or its derivatives), then the field equation
(\ref{eq:field2}) implies that $g_{\mu\nu}$ and/or $q_{\mu\nu}$
will accordingly acquire the corresponding property. However, as
$R_{\mu\nu}$ is constructed from the connection
$\Gamma^\alpha_{\beta\gamma}$ given by (\ref{connection}), it
contains the second-order derivatives of $q_{\mu\nu}$. So it is
difficult, if not impossible, to maintain the balance of the field
equation (\ref{eq:field1}) at these discontinuities if the
second-order derivatives of $q_{\mu\nu}$ do not exist. Thus, we
expect that the discontinuities in $T_{\mu\nu}$ (or its
derivatives) will give rise to similar discontinuities in
$g_{\mu\nu}$, but not in $q_{\mu\nu}$. On the other hand, in GR
the density profile of compact stars (or its spatial derivatives)
is in general discontinuous due to the presence of phase
transitions. Therefore, in the following discussion we consider
compact stars in EiBI gravity and study specifically how phase
transitions  could affect the equilibrium configuration and the
Ricci scalars derived respectively from  $g_{\mu\nu}$ and
$q_{\mu\nu}$.

%%%%%%%%%%%%%%%%%%%%%%%%%%%%%%%%%%%%%%%%%%%%%%%%%%%%%
\section{Static equilibrium of compact stars}\label{S3}
\label{S3}
%%%%%%%%%%%%%%%%%%%%%%%%%%%%%%%%%%%%%%%%%%%%%%%%%%%%%

%\subsection{Equilibrium conditions}
The structure of compact stars in EiBI theory has been studied by
Pani {\it et al.} \cite{Pani11,Pani12} and Sham {\it et al.}
\cite{Sham12}. Here we briefly review and follow the approach
developed in Ref. \cite{Sham12}. For a static and spherically symmetric
spacetime, the spacetime metric $g_{\mu\nu}$ and the auxiliary
metric $q_{\mu\nu}$ are taken to be
\begin{eqnarray}
g_{\mu\nu}dx^\mu dx^\nu &=& -e^{\phi (r)}dt^2+e^{\lambda (r)}dr^2
+f(r)d\Omega^2, \label{eq:metric_bkg}  \\
q_{\mu\nu}dx^\mu dx^\nu &=& -e^{\beta (r)}dt^2+e^{\alpha
(r)}dr^2+r^2d\Omega^2.\label{eq:metric_2}
\end{eqnarray}
 The compact star is made of  a perfect
fluid described by the standard stress-energy tensor
\begin{equation}
T^{\mu\nu} = (\epsilon + P) u^{\mu} u^{\nu} + P g^{\mu \nu} ,
\end{equation}
where $\epsilon$ and  $u^{\mu}$ are the energy density and
four-velocity of the fluid, respectively.

The field equations (\ref{eq:field1}) and (\ref{eq:field2}) lead
to a set of relations for the functions $\phi$, $\lambda$, $f$,
$\alpha$, and $\beta$ introduced in Eqs. (\ref{eq:metric_bkg}) and
(\ref{eq:metric_2}),
\begin{eqnarray}
{1\over \kappa}\left( 2+\frac{a}{b^3}-\frac{3}{a b}\right) &=&
\frac{2}{r^2}-\frac{2e^{-\alpha}}{r^2}+\frac{2e^{-\alpha}\alpha'}{r}
  ,
\label{eq:plus_static}  \\
{1\over \kappa}\left( \frac{1}{ab}+\frac{a}{b^3}-2\right) &=&
-\frac{2}{r^2}+\frac{2e^{-\alpha}}{r^2}+\frac{2e^{-\alpha}\beta'}{r}
, \label{eq:minus_static} \\
e^{\beta}&=&e^{\phi}b^3 a^{-1}, \\
e^{\alpha}&=&e^{\lambda}a b,
\label{eq:q_static}\\
f(r)&=&\frac{r^2}{ab} , \label{eq:F(r)}
\end{eqnarray}
where $a\equiv\sqrt{1+ 8\pi \kappa\epsilon}$ and $b\equiv\sqrt{1-
8\pi \kappa P}$, and hereafter primed quantities denote partial
derivatives with respect to $r$. Besides, the conservation of the
stress-energy tensor, Eq. (\ref{eq:matter_conserv}), gives another
relation,
\begin{equation}
\phi'=-\frac{2P'}{P+\epsilon}\label{eq:conserve_static}.
\end{equation}
Combining Eqs. (\ref{eq:plus_static}) \-- (\ref{eq:conserve_static}), one
can obtain two first-order differential equations governing the
structure of a compact star,
\begin{eqnarray}
\frac{d P}{d r} &=& -\Theta \left[\frac{2 }{ \epsilon + P
}+\frac{\kappa}{2} \left( \frac{3}{b^2} + \frac{1}{a^2 c_s^2}
\right) \right]^{-1}  \left[1-\frac{2m}{r}\right]^{-1},\nonumber \\
\label{eq:P'}
\\   \frac{d m}{d r} &=& {1\over 4 \kappa} \left( 2-\frac{3}{ab}
+\frac{a}{b^3} \right) r^2 ,   \label{eq:m'}
\end{eqnarray}
where
\begin{equation}\label{Q}
\Theta \equiv\left[ {1\over 2\kappa} \left( \frac{1}{a
b}+\frac{a}{b^3}-2 \right) r + \frac{2m}{r^2} \right],
\end{equation}
the speed of sound $c_s$ is calculated from the EOS by $c_s^2 =
dP/d\epsilon$, and the function $m(r)$ is defined by
\begin{equation}
e^{-\lambda}=\left( 1-\frac{2m}{r} \right) ab .
\label{eq:e_lambda}
\end{equation}
Equations (\ref{eq:P'}) and (\ref{eq:m'}) are analogous and in the
limit $\kappa \rightarrow 0$ reducible  to the well-known
Tolman-Oppenheimer-Volkov (TOV) equations in GR \cite{Oppenheimer,
Tolman}. With a given EOS $P=P(\epsilon)$ and suitable boundary
conditions (see below), these two equations completely determine
the hydrostatic equilibrium configuration of a compact star in
EiBI gravity. It follows directly from Eqs. (\ref{eq:m'}) and
(\ref{eq:e_lambda}) that $m(r=0)=0$, and $m(r) \geq 0$ increases
monotonically with $r$. Besides, it can be shown that $\Theta
>0 $ for positive $\kappa $ and $\Theta \approx 8 \pi P \{1+\pi
\kappa [(\epsilon-3P)^2/P +8\epsilon ]+{\cal O}(\kappa^2)\}
+{2m}/{r^2}$. As the typical value of $8 \pi |\kappa| \epsilon$ at
the center of a compact star is less than 0.4 \cite{Sham12}, for
specificity we assume in the present paper $\Theta
>0 $. In fact, we have numerically verified that $\Theta
>0 $ except for $P \ll \epsilon$.

The  boundary conditions supplementing Eqs. (\ref{eq:P'}) and
(\ref{eq:m'}) are as follows. First of all, the radius of the star
$R$ is as usual defined by the condition $P(R)=0$. Outside the
star where $r>R$, EiBI gravity is equivalent to GR, and
$g_{\mu\nu}$ is identical to the Schwarzschild metric (see, e.g.,
Refs. \cite{Weinberg72,Shapiro83}). As a consequence, the appropriate
boundary conditions at the stellar surface $r=R$ are
$\epsilon=P=0$, $a=b=1$, $ e^{-\alpha}= e^{-\lambda} = e^{\beta} =
e^{ \phi}=1 - 2M/R$, where $M \equiv m(R)$ is the mass of the
star.

The equilibrium configuration of compact stars has been studied in Refs. \cite{Pani11,Pani12,Sham12}. However, the compact stars considered in these references are all characterized by the smooth EOSs
$P=P(\epsilon)$ where $c_s^2 = dP/d\epsilon$ can be unambiguously
defined, and is continuous and nonvanishing except at the stellar
surface. Unlike the TOV equations in GR, the hydrostatic
equilibrium equations (\ref{eq:P'}) and (\ref{eq:m'}) demonstrate an explicit dependence on $c_s^2$. Therefore, it is reasonable to
expect that the behavior of $c_s^2$ could strongly affect the
equilibrium configuration of a compact star. In the following we
consider two different cases, namely, (i) $c_s^2=0$ (as in
first-order phase transitions) and (ii) $c_s^2$ is discontinuous
(as in both first- and second-order phase transitions).

\section{Stars with first-order phase transitions}\label{S4}
 For an EOS with a
first-order phase transition (see, e.g., Refs. \cite{Sotani01,PSR,1st}),
there is an interval in which the energy density increases while
the pressure remains the same, and consequently $c_s^2=0$. To
examine the consequence and other related problems of such
first-order phase transitions for compact stars in EiBI gravity,
we first consider cases with $c_s^2 \approx 0$ and expand Eq.
(\ref{eq:P'}) up to order $c_s^2$,
\begin{eqnarray}\label{1-1}
\frac{d P}{d r} & \approx &  -\frac{2 c_s^2 a^2\Theta}{\kappa}
\left[1-\frac{2m}{r}\right]^{-1} . \label{1-2}
\end{eqnarray}
It should be noted that $d\epsilon/dr=c_s^{-2}dP/dr$ is still
nonvanishing as $c_s^2 \rightarrow 0$. We study the implications
for the cases $\kappa
>0$ and $\kappa <0$ separately.

\begin{figure}[h!]
\includegraphics[scale=0.35,angle=0]{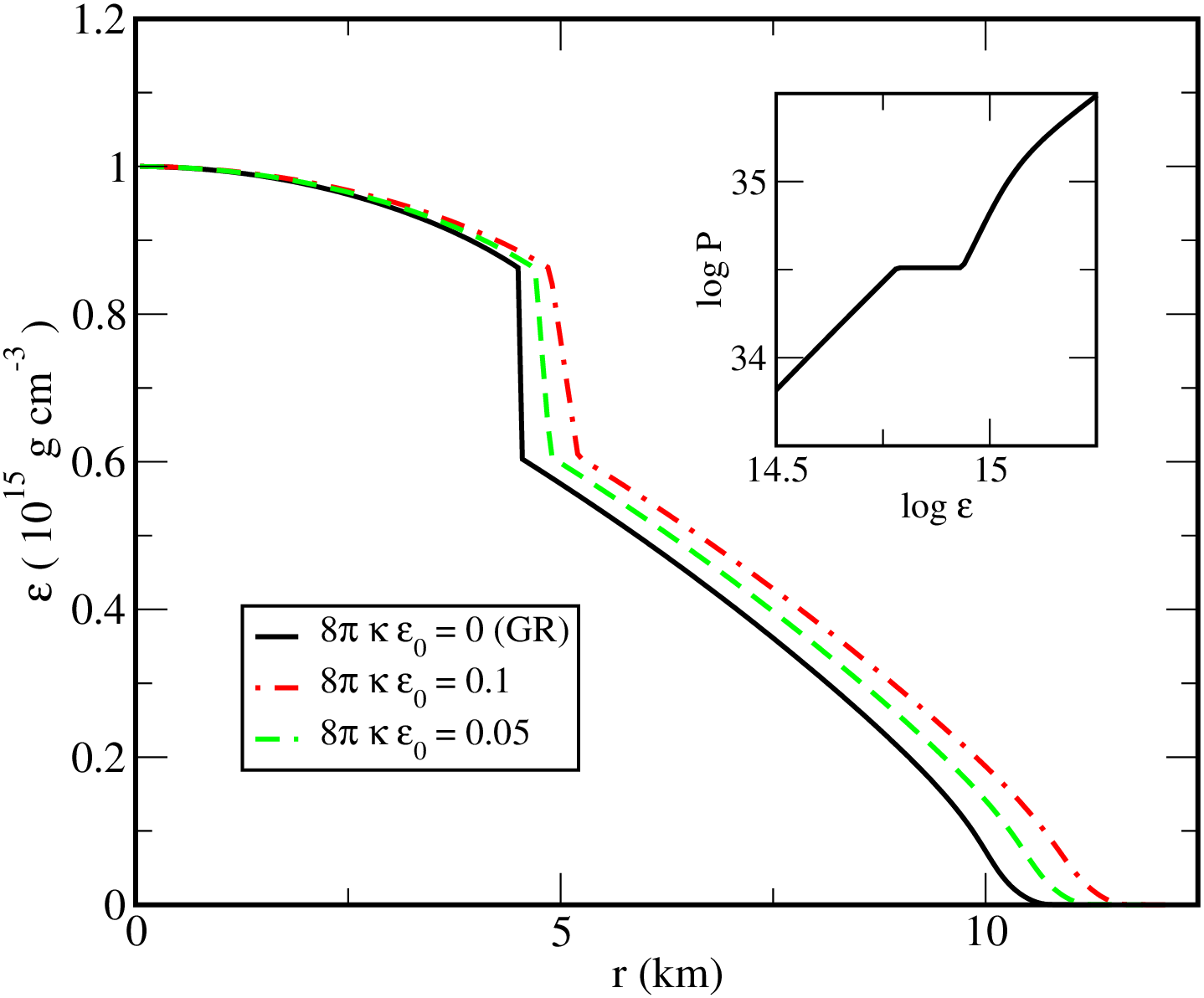}
\caption{The density profile of a compact star with an EOS with a
first-order phase transition as shown in the inset ($P$ and
$\epsilon$ are in cgs units)  is studied in both  GR (i.e. $\kappa
=0$) and EiBI gravity ($8\pi\kappa \epsilon_0 = 0.1, 0.05$, with
$\epsilon_0 \equiv 1\times10^{15} {\rm g}\, {\rm cm}^{-3}$). The
central density is also fixed at $\epsilon_0$
}\label{positive-first}
\end{figure}

First of all, for $\kappa > 0$, we can still get a normal energy
density profile in the region with $c_s^2 \approx 0$, where
$d\epsilon/dr$ is approximately equal to a negative constant  and
$dP/dr \approx 0^-$. As shown in Fig.~\ref{positive-first}, where a
first-order phase transition really takes place at a certain
pressure (the EOS is adapted from Ref.~\cite{1st}), there is a
finite region with constant $d\epsilon/dr$. Therefore the energy
density is now a continuous function. Physically, this corresponds to
a "mixed phase" region  (e.g., quark-nucleon mixed phase
\cite{Sotani01,PSR}) with constant pressure. The thickness of this
region is proportional to $\kappa$, and hence  vanishes in GR,
resulting in a discontinuity in the density profile (see
Fig.~\ref{positive-first}). The existence of a constant pressure
shell in compact stars built upon EiBI gravity is attributable to
the effective repulsive gravity inherent in EiBI gravity with
positive $\kappa$. Actually, the effective repulsion also leads to
the formation of pressureless stars, as suggested in Refs. 
\cite{Pani11,Pani12}.

On the other hand, if $\kappa < 0$ while $c_s^2 \approx 0$, both
$dP/dr$ and $d\epsilon/dr$ will become \textit{positive}, and in
the limit of $c_s^2 \rightarrow 0$, $dP/dr \rightarrow 0^+$ and
$d\epsilon/dr$ tends to a nonvanishing {\it positive} number.
Such behavior is in stark contrast to the case of GR where both
$dP/dr$ and $d\epsilon/dr$ are negative definite. If one starts
with a certain energy density at the center of a star and
integrates Eqs. (\ref{eq:P'}) and (\ref{eq:m'}) outward, one will find
that the energy density  drops initially and then rises again in
the density  region where a first-order phase transition occurs
and $c_s^2 = 0$. As a result, the energy density cannot drop to
zero and the surface of the star is undefined, implying that
equilibrium compact stars characterized by EOSs with first-order
phase transitions do not exist if $\kappa < 0$.

As a matter of fact, for stars with a sufficiently soft phase
where $c_s^2$ is small, we note a similar problem for EiBI
gravity with $\kappa < 0$. We focus on the last term of the modified
TOV equation (\ref{eq:P'}),
\begin{equation}
\frac{d P}{d r} \propto -\left[\frac{2 }{\epsilon +
P}+\frac{\kappa}{2}\left(\frac{3}{b^2} + \frac{1}{a^2
c_s^2}\right) \right]^{-1} .
\end{equation}
Given an equation of state, this term depends solely on
the energy density. When $\kappa = 0$, i.e., in the case of GR,
it is always negative. However, when $\kappa < 0$ and
$c_s^2$ is small, this term may blow up at certain values of
energy density regions, and consequently the modified TOV equation
cannot be solved. This again implies that for $\kappa < 0$, compact stars
with soft EOSs may not be able to support an equilibrium state.

The predicament encountered in solving Eq. (\ref{eq:P'}) in the
presence of first-order phase transitions can also be understood
in terms of the concept of the apparent EOS proposed by Delsate and
Steinhoff \cite{Delsate_12}, who have shown that EiBI gravity can
be recast as ordinary GR provided that the physical EOS is
replaced with an apparent EOS $P_q(\epsilon_q)$, where for perfect
fluids the apparent pressure $P_q$ and density $\epsilon_q$ are
given respectively by $P_q = \tau P +{\cal P}$, $\epsilon_q  =
\tau \epsilon -\cal{P}$, with $\tau =[(1+8 \pi\kappa \epsilon)(1-
8 \pi \kappa P)^3]^{-1/2}$ and ${\cal P} = [\tau - 1 - \kappa \tau
(3P-\epsilon)/2]/\kappa$. In first-order phase transitions, where
$d P/d\epsilon= 0$,
\begin{equation}\label{}
\frac{d P_q}{d\epsilon_q}   = \frac{8 \pi\kappa
(P+\epsilon)}{a^2+3b^2}.
\end{equation}
Therefore,  for negative $\kappa$, $d P_q/d\epsilon_q < 0$. In
this case, despite the fact that the TOV equations in GR guarantee
that $P_q$ decreases monotonically towards the stellar surface,
the apparent density $\epsilon_q$ increases instead in the region
where a first-order phase transition takes place. Hence, by the
same token, it is also impossible to construct the equilibrium
configuration of a compact star governed by the TOV equations and
the apparent EOS.

\section{Singular behavior of Ricci scalar}\label{S5}
In the above discussion, we see that the equilibrium configuration of
compact stars can generally be  constructed in EiBI gravity with
positive $\kappa$ even in the presence of phase transitions. In
particular, the energy density $\epsilon$ is always a continuous
function of the radial coordinate $r$. However, as noted in
Ref.~\cite{Pani_sing} and will be shown in the present paper, the
associated Ricci scalar  still acquires certain singular behaviors
(e.g., discontinuities and divergence) due to higher derivatives
of $\epsilon$.

In EiBI gravity, the Ricci scalar can be constructed using the
metrics $q_{\mu\nu}$ and $g_{\mu\nu}$, which are respectively
given by
\begin{widetext}
\begin{eqnarray}
R_q =&&
-\frac{1}{2r^2e^\alpha}\left[-\beta'\alpha'r^2+2\beta''r^2+\beta'^2r^2+4r\beta'
-4r\alpha'-4e^\alpha+4\right],\label{Rq} \\
R_g =&& -\frac{1}{2 e^\lambda}\left[-\phi'\lambda' + 2\phi'' +
\phi'^2+2(\phi'-\lambda')\frac{f'}{f}
-\left(\frac{f'}{f}\right)^2+4\frac{f''}{f}-4\frac{e^\lambda}{f}
\right]. \label{Rg}
\end{eqnarray}
\end{widetext}

From the expressions of $R_q$ and $R_g$, Eqs. (\ref{Rq}) and
(\ref{Rg}), one can show that $R_q$ depends on $c_s^2$ while $R_g$
depends on both $c_s^2$ and $(c_s^2)'$. In the presence of phase
transitions of the first or second order, $c_s^2$ is in general
discontinuous at the transition point (see, e.g., the inset in
Fig.~1). Hence, it is likely that these Ricci scalars could become
discontinuous or even blow up at radius $r_t$ where the phase
transition occurs. However, it is readily shown that $\alpha$,
$\alpha'$, $\beta$, $\beta'$ and $\beta''$ are indeed continuous
functions of $r$ even in the presence of phase transitions. Hence,
$R_q$ is still well-behaved even in the vicinity of $r=r_t$, which
agrees with our intuition as mentioned in Sec.~\ref{S2}. Besides,
as shown in Ref.~\cite{Delsate_12} and discussed in the last
section, $R_q$ is indeed the standard Ricci scalar in GR for a
star constructed with the apparent EOS. Therefore, its regularity
is also expected.

The situation is completely different for $R_g$. It contains a
$\delta$-function singularity given by
\begin{equation}
R_g = \frac{\beta'}{e^{\lambda}}
[A_1(A_2+2A_3)]_-\delta(r-r_t)+...~,\label{eq:delta}
\end{equation}
where the notation $[F]_-\equiv F(r=r_t^+)-F(r=r_t^-)$  for a
physical quantity $F$ is introduced,
\begin{eqnarray}
A_1&=&\left\lbrace\left(\frac{4}{a^2-b^2}+\frac{3}{b^2}\right)c_s^2+\frac{1}{a^2}\right\rbrace^{-1},\label{eq:A1}\\
A_2&=&\frac{3c_s^2}{b^2}+\frac{1}{a^2},\label{eq:A2}\\
A_3&=&\frac{c_s^2}{b^2}-\frac{1}{a^2},\label{eq:A3}
\end{eqnarray}
and other finite terms have been suppressed. In addition to the
$\delta$-function singularity, $R_g$ is also discontinuous there,
with a jump given by
\begin{widetext}
\begin{eqnarray}
[R_g]_-&=&-\frac{1}{2e^\lambda}\left\lbrace(-\frac{4}{r}\beta'-2\beta''+\alpha'\beta')[A_1(A_2+2A_3)]_-
-\beta'^2[A_1(2A_2+A_3)]_-+\beta'^2[A_1^2(A_2^2+A_2A_3+A_3^2)]_- \right. \nonumber \\
&&-\left.2\beta'\left[\frac{d}{dr}A_1(A_2+2A_3)\right]_-\right\rbrace.
\label{gap}
\end{eqnarray}
\end{widetext}
%\begin{widetext}
%\begin{eqnarray}
%[R_g]_-&=&-\frac{1}{2e^\lambda}\left\lbrace(-\frac{4}{r}\beta'-2\beta''+\alpha$
%&&-\beta'^2[A_1(2A_2+A_3)]_-+\beta'^2[A_1^2(A_2^2+A_2A_3+A_3^2)]_-\nonumber\\
%&&-\left.2\beta'\left[\frac{d}{dr}A_1(A_2+2A_3)\right]_-\right\rbrace,
%\label{gap}
%\end{eqnarray}
%\end{widetext}
For reference purpose, the detailed forms of $A_i'$ ($i=1,2,3$)
are given below:
\begin{eqnarray}
A_1' &=&
-A_1^2\left\{\frac{3a^2+b^2}{b^2(a^2-b^2)}(c_s^2)' \right. \nonumber \\
&&+\left.\beta'A_1
\left(\frac{8(1+c_s^2)c_s^2}{(a^2-b^2)^2}-\frac{c_s^4}{b^4}-\frac{2}{a^4}\right)\right\},\\
A_2'&=&\frac{3}{b^2}(c_s^2)'-2A_1\beta'\left(\frac{3c_s^4}{b^4}-\frac{1}{a^4}\right),\\
A_3'&=&
\frac{1}{b^2}(c_s^2)'-2A_1\beta'\left(\frac{c_s^2}{b^4}+\frac{1}{a^4}\right).
\end{eqnarray}

In general, whereas the $\delta$-function singularity in $R_g$ is
directly proportional to the discontinuity in $c_s^2$, the jump
$[R_g]_-$ depends on both $[c_s^2]_-$ and $[(c_s^2)']_-$. Thus,
for compact stars where first- or second-order phase transitions
take place, the Ricci scalar $R_g$ is singular in the sense that
it contains a $\delta$-function-type divergence and spatial
discontinuity. As is well known, the Ricci scalar in GR is
proportional $- \epsilon + 3P$ and is always bounded. The
$\delta$-function-type divergence in $R_g$ discovered here  is a
distinctive feature of EiBI gravity.

\begin{figure}[h!]
\centering
\subfigure[]{\includegraphics*[width=9cm,angle=0]{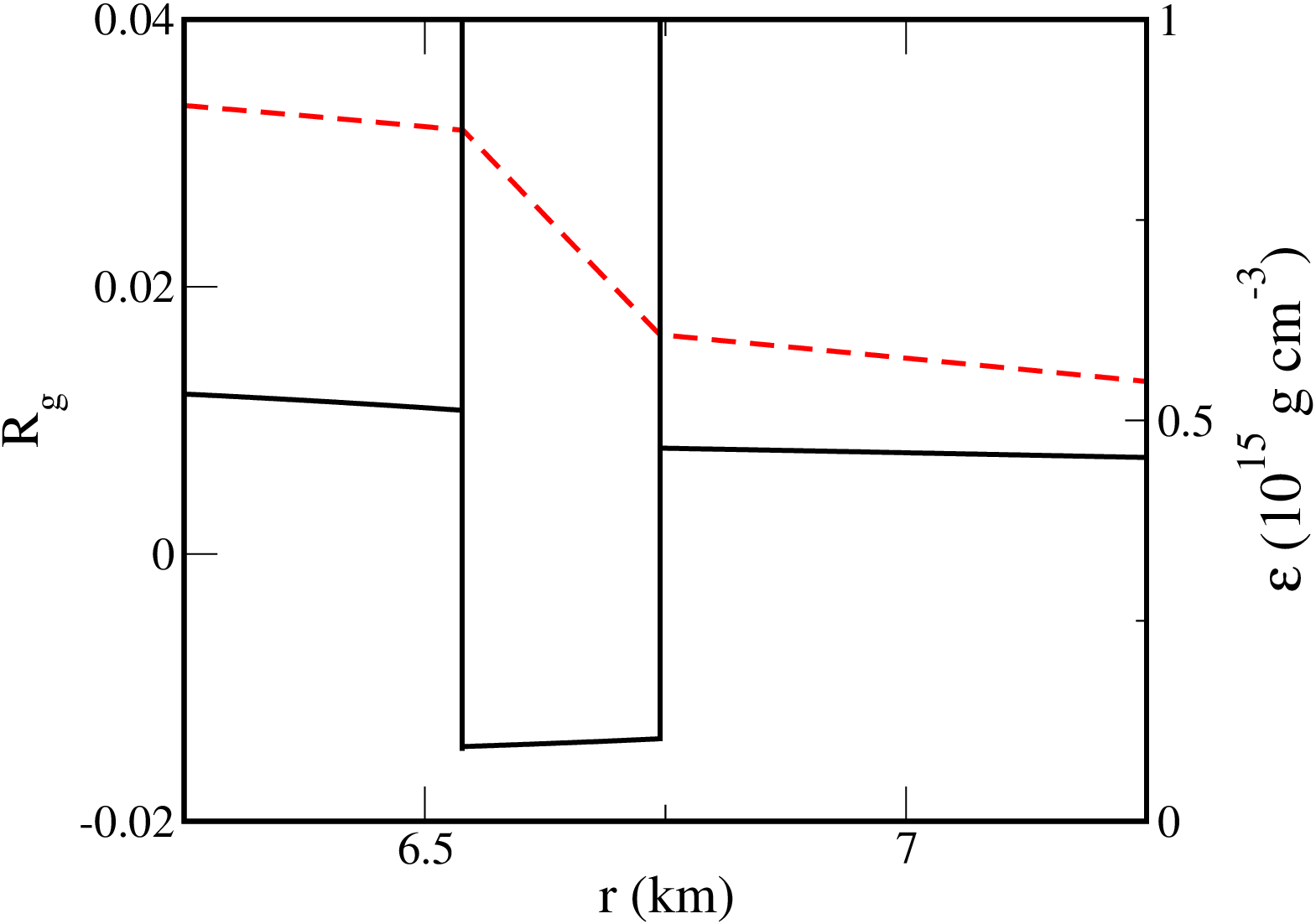}}
\subfigure[]{\includegraphics*[width=9cm,angle=0]{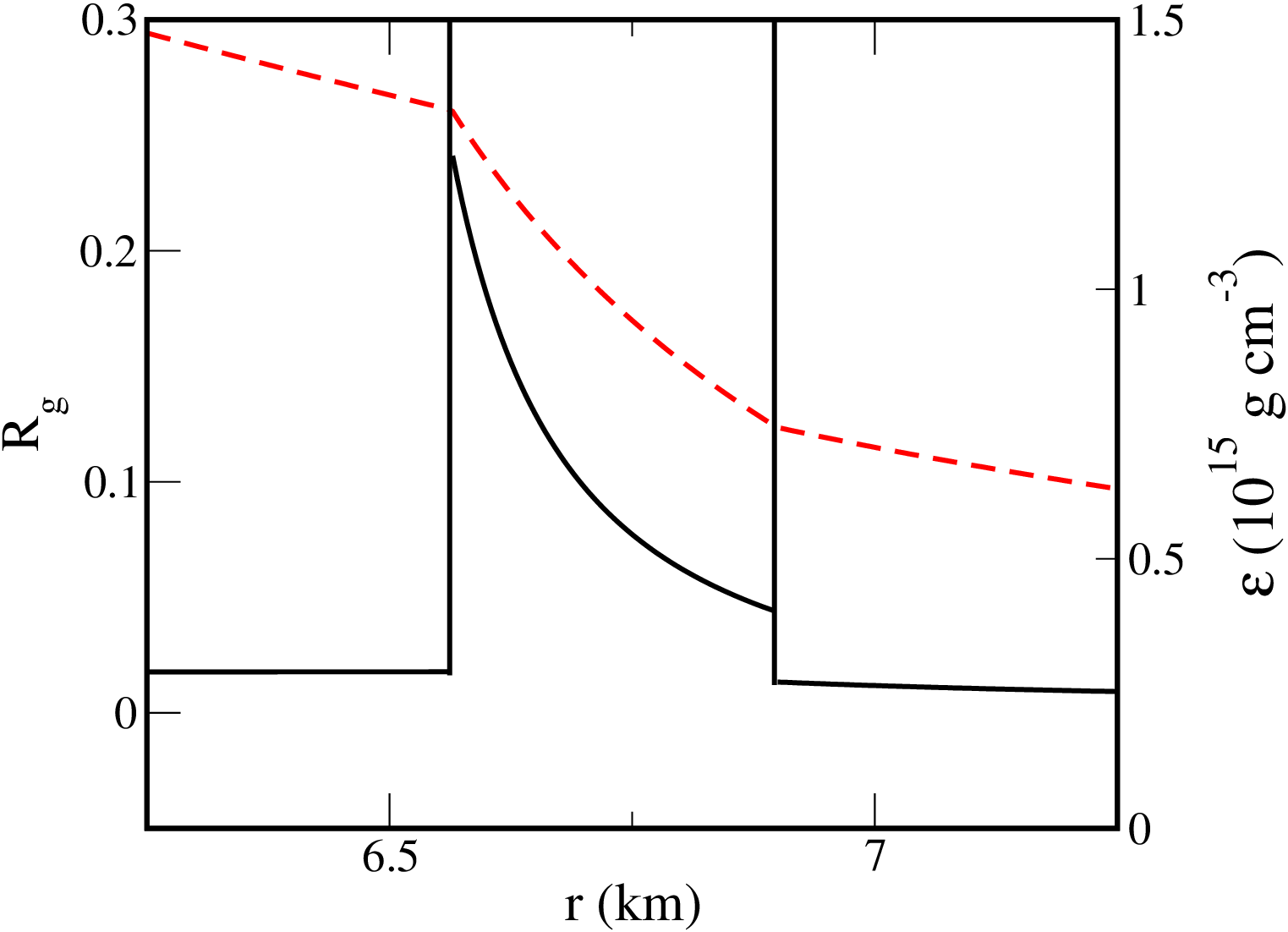}}
\caption{The Ricci scalar $R_g$ (solid line, left-scale) and
energy density $\epsilon$ (dashed line, right-scale) of a compact
star where a first- (upper panel, central density $= 1.2\epsilon_0$
) or second-order (lower panel, central density $= 3 \epsilon_0$)
phase transition occurs are plotted against the radius $r$.
$8\pi\kappa\epsilon_0$ is equal to $0.1$ for both cases. For
$R_g$, $\delta$-function singularities (as indicated by the
vertical lines in the figures) emerge at the places where $c_s^2$
is discontinuous.} \label{fig:R_g}
\end{figure}

 We have numerically verified the said behavior of
$R_q$ and $R_g$.   For EOSs with phase transitions, either of
first- or second-order (the EOSs are adapted from Refs.~\cite{1st,Banik03}, respectively), $R_q$ is found to be a
continuous function, while $R_g$ (see Fig.~\ref{fig:R_g}) is
discontinuous and indeed blows up at the place where there is a
jump in $c_s^2$. The discontinuity in $R_g$ obtained numerically
in fact agrees with the expression given by Eq. (\ref{gap}).

\section{Discussion}\label{S6}

In this paper, extending our previous work \cite{Sham12}, we have
studied compact stars in EiBI gravity using realistic EOS models
with phase transitions and discovered several anomalies in the
behavior of such stars. For EOSs with first-order phase
transitions, compact stars in EiBI gravity with $\kappa > 0$
exhibit an anomalous "mixed phase" region where  $c_s^2 =0$,
the pressure is a constant and the energy density $\epsilon$ is still
a continuous function of the radius $r$. This is in contrast to
the situation in GR where $\epsilon$ is discontinuous at the
transition point. It is the effective pressure inherent in EiBI
gravity with positive $\kappa$ that leads to the existence of the
constant pressure region. On the other hand, for $\kappa < 0$, the
equilibrium configuration for compact stars with first-order phase
transitions (or  soft enough EOSs)  cannot be constructed because
$\epsilon$ never vanishes and hence it is not possible to fulfill
the boundary conditions ($\epsilon=P=0$) at the stellar surface in
order to match the interior solution smoothly to the Schwarzschild
spacetime.

As a side remark, in general for $\kappa \ne 0$,  we find that one
cannot construct the simplest  quark star model, described by the
MIT bag model EOS, since $\epsilon$ is finite when $P=0$ for this
EOS. Note that this does not pose a problem in GR because one only
requires $P=0$ at the surface in GR.

In GR, one can always construct a compact star model by solving
the TOV equation with a prescribed EOS. The EOS model is required
to satisfy only a few general conditions which are already imposed
by the microscopic theory (e.g., the stability condition
$dP/d\epsilon \ge 0$ and the causality limit $c_s<c$). On the
contrary, EiBI gravity appears to put a more severe restriction on
what kind of EOS one can use to build a stellar model. It is true
that whether compact stars in nature exhibit a phase transition in
their interiors is still a matter of debate. Nevertheless, it
seems unreasonable that the underlying gravitational theory would
put a constraint on which microscopic EOS model one can use to
construct a theoretical stellar model in the first place. Even in
the situation where a compact stars with a first-order phase
transition can be constructed in EiBI theory (i.e., when $\kappa >
0$), we still need to face the $\delta$-function singularity of
the Ricci scalar $R_g$ as we have shown in Sec.~\ref{S5}.

Our discovery reported here complements the recent work of Pani
and Sotiriou \cite{Pani_sing} in which they demonstrated the
singularity of $R_g$ at the surface of a polytropic sphere for any
$\Gamma > 3/2$. In fact, the singular behaviors of $R_g$
discussed in the present paper and
Ref.~\cite{Pani_sing}  stem from a common physical origin, namely
the emergence of the first- and second-order derivatives of the
pressure in the Ricci scalar $R_g$. These derivatives in the
matter field arise as a consequence of the elimination of the
auxiliary metric $q_{\mu \nu}$ defined in Eq. (\ref{connection}). (See
Ref.~\cite{Pani_sing} for a detailed discussion. ) They are absent
in the GR case and lead to enigmatic singularities in the Ricci
scalar, as mentioned above.

In conclusion, EiBI gravity is appealing because it can avoid some
of the singularities that plague GR by introducing nonlinear
coupling between matter and gravity. However, it is also the same
nonlinear coupling that leads to the anomalies of compact stars with
phase transitions, as we have found in this paper. This renders the
viability of EiBI gravity questionable.

\begin{acknowledgments}We thank an anonymous referee for drawing
our attention to the apparent EOS approach to EiBI gravity
proposed in Ref.~\cite{Delsate_12}. \end{acknowledgments}

\bibliographystyle{apsrev}
%\bibliography{EibI-ref}

\end{document}